\documentclass[showpacs,aps,prd,preprint,nofootinbib,showkeys,unsortedaddress,raggedbottom]{revtex4-1}
\pdfoutput=1

\usepackage{bm}
\usepackage{amsmath}
\usepackage{graphicx}
\usepackage{subfigure}
\usepackage[usenames,dvipsnames]{color}
\definecolor{darkblue}{RGB}{0,0,196}
\usepackage[colorlinks=true,linkcolor=darkblue,citecolor=darkblue,urlcolor=darkblue]{hyperref}
\usepackage{slashed}

\usepackage{setspace}
\usepackage{footmisc}
\usepackage[makeroom]{cancel}

\def\be{\begin{equation}}
\def\ee{\end{equation}}
\def\ba{\begin{eqnarray}}
\def\ea{\end{eqnarray}}

\begin{document}

\title{Far-from-equilibrium attractors with a realistic non-conformal equation of state}

\author{Mubarak Alqahtani} 
\affiliation{Department of Physics, College of Science, 
Imam Abdulrahman Bin Faisal University, Dammam 31441, Saudi Arabia }

\begin{abstract}
Using anisotropic hydrodynamics, we examine the existence of early-time attractors of non-conformal systems undergoing Bjorken expansion.  In the case of a constant mass, we find that the evolution of the scaled longitudinal pressure is insensitive to variations of initial conditions converging onto an early-time universal curve and eventually merging with the late-time Navier-Stokes attractor (the hydrodynamic attractor). On the other hand, the bulk and the shear viscous corrections do not show an early-time attractor behavior. These results are consistent with previous studies considering a constant mass. When a realistic equation of state is included in the dynamics with a thermal mass, we demonstrate for the first time the absence of strict late-time universal attractors. However, a semi-universal feature of the evolution at very late times remains. 
\end{abstract}

\date{\today}



\maketitle

\section{Introduction}
\label{sec:intro}

Relativistic hydrodynamics has been a successful tool in modeling relativistic heavy-ion collisions~\cite{nucl-th/0305084,0902.3663,1301.5893,1605.08694,Jeon:2016uym,1707.02282,Romatschke:2017ejr}. Throughout the years, different dissipative hydrodynamics approaches have been used to describe final-state observables~\cite{Huovinen:2001cy,Romatschke:2007mq,0911.2397,Niemi:2011ix,1009.3244,1209.6330,Ryu:2015vwa,1409.8164,Alqahtani:2017jwl,1711.08499,1901.01319} (For a recent review, see Ref.~\cite{2010.12377}). This success in predicting and describing the experimental results was not limited to large systems such as Pb-Pb and Au-Au, but was also seen in small systems such as p-Pb~\cite{1312.6555,1502.04745,1609.02590,1701.07145,1705.03177,1801.03477}. This applicability in small systems seems to contradict the assumption that hydrodynamics applicability is limited to systems with near thermal equilibrium which is not expected in small systems at all.  This puzzle triggered great interest in understanding the domain of applicability of hydrodynamics~\cite{1704.08699}.

 The applicability of hydrodynamics could be explained by the existence of the attractor solutions in far-from-equilibrium systems.  In such systems, the evolution of solutions-irrespective of their initial conditions-  converge quickly to a universal solution (the attractor) at very early times. This universality seems to be a property of hydrodynamics where different solutions lose information about their initial conditions. There have been many interesting works where attractor solutions have been found and examined using different approaches and symmetries see e.g., \cite{1503.07514,1512.05347,1603.05344,1511.06358,1704.08699,1708.06255,1708.01921,1709.06644,1809.01200,1903.03145,2004.05195}. We note that almost all of the attractor studies cited above have focused on simple setups where the systems under consideration were conformal. Recent reviews about attractors can be found in~\cite{1707.02282,2109.15081}.

 Recently, the effect of nonconformality on the attractors for different approaches has been studied in~\cite{ 1710.03234,1710.07095,1907.08101, 1907.07965,2107.05500,2107.10248,2208.02750,2210.00658}. In Ref.~\cite{2107.10248} specifically, a modified anisotropic hydrodynamics approach is introduced and compared to the exact kinetic theory solutions. The agreement found using this approach to the kinetic theory solutions is excellent,  as pointed out by the authors, especially at early times and for the largest possible initial negative bulk pressures. Such an agreement between anisotropic hydrodynamics and exact solutions motivates us to study the existence of early-time attractors of anisotropic hydrodynamics using a realistic non-conformal equation of state for quasiparticles.

This work is an attempt to study the effect of using a realistic non-conformal equation of state on hydrodynamic attractor solutions. To do so, we will use anisotropic hydrodynamics approach of systems undergoing Bjorken expansion of quasiparticles having a thermal mass $m(T)$~\cite{1509.02913}. This model is called quasiparticle anisotropic hydrodynamics and has shown a good phenomenological agreement with experimental results in the 3+1D case, see e.g.~\cite{1703.05808,1705.10191,2007.04209,2008.07657,2209.10894}.   In this work, however, we limit ourselves to the simplest case of 0+1D and try to examine if early-time attractors survive when a realistic non-conformal equation of state is assumed in the dynamics.

The structure of the manuscript is as follows. In Sec.~\ref{sec:aHydroQP}, we introduce anisotropic hydrodynamics generally, then obtain the dynamical equations needed for the system's evolution in the quasiparticle approach. In Sec.~\ref{sec:results}, we show our results of the early-time attractors for the scaled longitudinal pressure and the shear stress pressure. For comparison, we also show the results of systems with a constant mass using anisotropic hydrodynamics as well. Unlike the constant mass case, where strict early-time attractors exist for the scaled longitudinal pressure, by using a realistic equation of state, we find a semi-universal attractor at very late times. Conclusions and a future outlook are summarized in Sec.~\ref{sec:conclusions}.


\section{Anisotropic hydrodynamics}
\label{sec:aHydroQP}

\subsection{3+1D anisotropic hydrodynamics}
\label{subsec:aHydroQP}

The anisotropic hydrodynamics (aHydro) approach is motivated by the fact that the quark-gluon plasma (QGP) dynamics is highly momentum-space anisotropic ~\cite{1007.0130,1007.0889} (see Ref.~\cite{1410.5786} for an introduction of aHydro). In this framework, the one-particle distribution function is assumed to be momentum-space anisotropic in the local-rest frame (LRF)~\cite{1405.1355}
\be
f_{\rm LRF}(x,p) =  f_{\rm eq}\!\left(\frac{1}{\lambda}\sqrt{\sum_i \frac{p_i^2}{\alpha_i^2} + m^2}\right) ,
\label{eq:fform}
\ee
with $\alpha_i$ being the anisotropy parameters ($i \in \{x,y,z\}$) and $\lambda$ being a parameter that is identified with the temperature in the isotropic equilibrium limit. In the limit where $\alpha_i=1$ and $\lambda=T$, one recovers the isotropic distribution function and in the case of Boltzmann statistics, $f_{\rm eq}=\exp(-E/T)$.

From kinetic theory, the distribution function of a gas of particles having mass $m(T)$ obeys the Boltzmann equation~\cite{1108.5561,1509.02913}
\be
p^\mu \partial_\mu f+\frac{1}{2}\partial_i m^2\partial^i_{(p)} f=-\mathcal{C}[f]\,,
\label{eq:boltzqp}
\ee
with $f$ being the assumed single-particle distribution function and $\mathcal{C}[f]$ is the collisional kernel. In this work, we assume $f $ as been given in Eq.~(\ref{eq:fform}) and $\mathcal{C}[f]$ to be in the relaxation-time approximation given by  $\mathcal{C}[f]=p^\mu u_\mu(f-f_{\rm eq})/\tau_{\rm eq}$ where $\tau_{\rm eq}$ is position dependent and given by
\be 
\tau_{\rm eq}(T)=\frac{15 \bar{\eta}}{\kappa(\hat{m}_{\rm eq})T}\bigg(1+\frac{{\cal E}_{\rm eq}(T)}{{\cal P}_{\rm eq}(T)}\bigg) \, ,
\label{eq:teq}
\ee
where $\kappa$ can be expressed in terms of modified Bessel functions of the second kind and modified Struve functions as defined in Ref.~\cite{1509.02913} where  $\hat{m}_{\rm eq}=m/T$ . Moreover, $\bar{\eta}$ is shear viscosity to entropy density ratio $\eta/{\cal S}_{\rm eq}$, which is held constant during the evolution of the system.  ${\cal E}_{\rm eq}$ and ${\cal P}_{\rm eq}$ are the equilibrium energy density and pressure, respectively.We note that in the conformal limit, Eq.~(\ref{eq:teq}) becomes
\be 
\tau_{\rm eq}(T)=\frac{5\eta}{4{\cal P}_{\rm eq}}= \frac{5\bar{\eta}}{T}\,.
\label{eq:teqconformal}
\ee
The dynamical equations can be obtained by taking the lower moments of the Boltzmann equation, Eq.~(\ref{eq:boltzqp})
\ba
\partial_\mu J^\mu &=& -\int dP \, {\cal C}[f]\, , \label{eq:J-conservation} \\
\partial_\mu T^{\mu\nu}&=&-\int dP \, p^\nu {\cal C}[f]\, , \label{eq:T-conservation} \\
\partial_\mu {\cal I}^{\mu\nu\lambda}- J^{(\nu} \partial^{\lambda)} m^2 &=&-\int dP \, p^\nu p^\lambda{\cal C}[f]\, \label{eq:I-conservation},
\ea 
which are the zeroth, first, and second moments of the Boltzmann equation, respectively.  $J^\mu$ is the particle four-current, $T^{\mu\nu}$ is the energy-momentum tensor, and ${\cal I}^{\mu\nu\lambda}$ is a rank-three tensor. They are given by
\ba
J^\mu &\equiv& \int dP \, p^\mu f(x,p)\, , \label{eq:J-int} \\
T^{\mu\nu}&\equiv& \int dP \, p^\mu p^\nu f(x,p)+B g^{\mu\nu}, \label{eq:T-int}\\
{\cal I}^{\mu\nu\lambda} &\equiv& \int dP \, p^\mu p^\nu p^\lambda  f(x,p) \, ,
\label{eq:I-int}
\ea
where $dP$ is the Lorentz invariant momentum-space integration measure given by $\tilde{N} \frac{d^3p}{E} $ with  $\tilde{N} \equiv N_{\rm dof}/(2\pi)^3$  where $N_{\rm dof}$ is the number of degrees of freedom. 

We note here that a background contribution $B$ is added to the definition of the energy-momentum tensor as shown in Eq.~(\ref{eq:T-int}) to ensure thermodynamic consistency when a thermal mass is considered in the dynamics. The basics of this approach which is called quasiparticle anisotropic hydrodynamics are explained in detail in Ref.~\cite{1509.02913}. The space-time dependence of $B$ can be obtained by the following partial differential equation relating $B$ and the thermal mass
\be
\partial_\mu B = -\frac{1}{2} \partial_\mu m^2 \int dP  f(x,p)\,.
\label{eq:BM-matching}
\ee
On the other hand, the thermal mass $m(T)$ is obtained by tuning to the equation of state (EoS) from lattice QCD calculations \cite{1007.2580}. These results provide an analytic parameterization of the interaction measure (trace anomaly) where the energy density and the pressure could be obtained ($I_{\rm eq}={\cal E}_{\rm eq}-3{\cal P}_{\rm eq}$). Once the equilibrium energy density and pressure are found one may use the following thermodynamic identity to find $m(T )$ as outlined in Ref.~\cite{1509.02913}
\be
{\cal E} _{\rm eq}+ {\cal P}_{\rm eq}=T {\cal S}_{\rm eq} = 4 \pi \tilde{N} T^4 \, \hat{m}_{\rm eq}^3 K_3\left( \hat{m}_{\rm eq}\right) ,
\ee

For a quasiparticle gas in an isotropic equilibrium state, the energy density and the pressure, assuming Boltzmann distribution, are given by
\ba
{\cal E}_{\rm eq}(T,m) &=& 4 \pi \tilde{N} T^4 \, \hat{m}_{\rm eq}^2
 \Big[ 3 K_{2}\left( \hat{m}_{\rm eq} \right) + \hat{m}_{\rm eq} K_{1} \left( \hat{m}_{\rm eq} \right) \Big]+B_{\rm eq} \, , 
\label{eq:Eeq} \\
 {\cal P}_{\rm eq}(T,m) &=& 4 \pi \tilde{N} T^4 \, \hat{m}_{\rm eq}^2 K_2\left( \hat{m}_{\rm eq}\right)-B_{\rm eq} \, ,
\label{eq:Peq}
\ea
with $\hat{m}_{\rm eq}=m/T$. 
\subsection{0+1D quasiparticle anisotropic hydrodynamics}
\label{subsec:aHydroQP0+1}
In this work, we consider boost-invariant systems (0+1D) where the energy density, transverse pressure, and longitudinal pressure are given by~\cite{1509.02913}
\ba
{\cal E} &=& \tilde{{\cal H}}_3({\boldsymbol\alpha},\hat{m}) \, \lambda^4+B \, ,\nonumber \\
{\cal P}_T &=& \tilde{{\cal H}}_{3T}({\boldsymbol\alpha},\hat{m}) \, \lambda^4-B \, ,\nonumber \\
{\cal P}_L &=& \tilde{{\cal H}}_{3L}({\boldsymbol\alpha},\hat{m}) \, \lambda^4-B \,.
\label{eq:E-P-trans}
\ea
In 0+1D, the space-time dependence of $B$ can be obtained from Eq.~(\ref{eq:BM-matching}) and can be written as
\be
\partial_\tau B = - \frac{\lambda^2 }{2} \tilde{\cal H}_{3B}({\boldsymbol\alpha},\hat{m}) \, \partial_\tau m^2  \, .
\label{eq:B}
\ee
The various  ${\cal{H}}$-functions appearing above are 3D integrals that can be done numerically. Their exact definitions can be found in Ref.~\cite{1509.02913}. As an example, $\tilde{\cal H}_3$ is defined as
\be
\tilde{\cal H}_3({\boldsymbol\alpha},\hat{m}) \equiv  2 \pi \tilde{N} \alpha_x^4
\int_0^\infty d\hat{p} \, \hat{p}^3  f\!\left(\!\sqrt{\hat{p}^2 + \hat{m}^2}\right) {\cal H}_2\!\left(\frac{\alpha_z}{\alpha_x},\frac{\hat{m}}{\alpha_x\hat{p}} \right) ,
\label{eq:h3tilde}
\ee
where
\be
 {\cal H}_2(y,z)
= \frac{y}{\sqrt{y^2-1}} \left[ (z^2+1)
\tanh^{-1} \sqrt{\frac{y^2-1}{y^2+z^2}} + \sqrt{(y^2-1)(y^2+z^2)} \, \right] ,\\
\label{eq:H2}
\ee

Here, we have four variables,  $\alpha_x$, $\alpha_y$, $\lambda$, and $T$, and we need four equations to find their temporal dependence which are solely obtained from the first and second moments. The dynamical equations of a system undergoing boost-invariant expansion according to Bjorken,  0+1D in the RTA approximation,  are derived in detail in Ref.~\cite{1509.02913}, we list them here for reference,
\ba
&& 4 \tilde{\cal H}_3 \partial_\tau\log\lambda+\tilde{\Omega}_m\partial_\tau\log \hat{m} +\tilde{\Omega}_L\partial_\tau\log\alpha_z
+\tilde{\Omega}_T\partial_\tau\log\alpha_x^2 +\frac{\partial_\tau B}{\lambda^4}+\frac{\tilde{\Omega}_L}{\tau} = 0 \, , \hspace{1cm} \label{eq:final1-thermal} \\
&& 4\partial_\tau\log\alpha_x+\partial_\tau\log\alpha_z+5\partial_\tau\log\lambda
+ \partial_\tau \log\!\left(\hat{m}^3 K_3(\hat{m})\right)
+\frac{1}{\tau}
\nonumber \\ && \hspace{8cm} 
= \frac{1}{\tau_{\rm eq}}\left[\frac{1}{\alpha_x^4\alpha_z}\Big(\frac{T}{\lambda}\Big)^2\frac{K_3(\hat{m}_{\rm eq})}{K_3(\hat{m})}-1\right] \,, 
\label{eq:final2xx-thermal} 
\\
&& 2\partial_\tau\log\alpha_x+3\partial_\tau\log\alpha_z+5\partial_\tau\log\lambda
+ \partial_\tau \log\!\left(\hat{m}^3 K_3(\hat{m})\right)
+\frac{3}{\tau} \nonumber \\ && \hspace{8cm} 
= \frac{1}{\tau_{\rm eq}}\left[\frac{1}{\alpha_x^2\alpha_z^3}\Big(\frac{T}{\lambda}\Big)^2\frac{K_3(\hat{m}_{\rm eq})}{K_3(\hat{m})}-1\right] \,,
\label{eq:final2zz-thermal}
\\
&&4 \tilde{\cal H}_{3,\rm eq} \partial_\tau\log T +\tilde{\Omega}_{m,\rm eq}\partial_\tau\log \hat{m}_{\rm eq}
+\frac{\tilde{\Omega}_L}{\tau}\Big(\frac{\lambda}{T}\Big)^4+\frac{\partial_\tau B}{T^4}=0\,.
\label{eq:final-matching-thermal}
\ea
where $\tilde{\Omega}_T$, $\tilde{\Omega}_L$, and $\tilde{\Omega}_m$ are functions of $\tilde{\cal H}$ functions defined in App.~B in Ref.~\cite{1509.02913}. In the case, where $m$ is taken to be constant i.e. $\partial_\tau m=0$ and $B=0$, the dynamical equations are derived similarly in Ref.~\cite{1405.1355}.

\begin{figure}[t!]
\centerline{
\hspace{-1.5mm}
\includegraphics[width=1\linewidth]{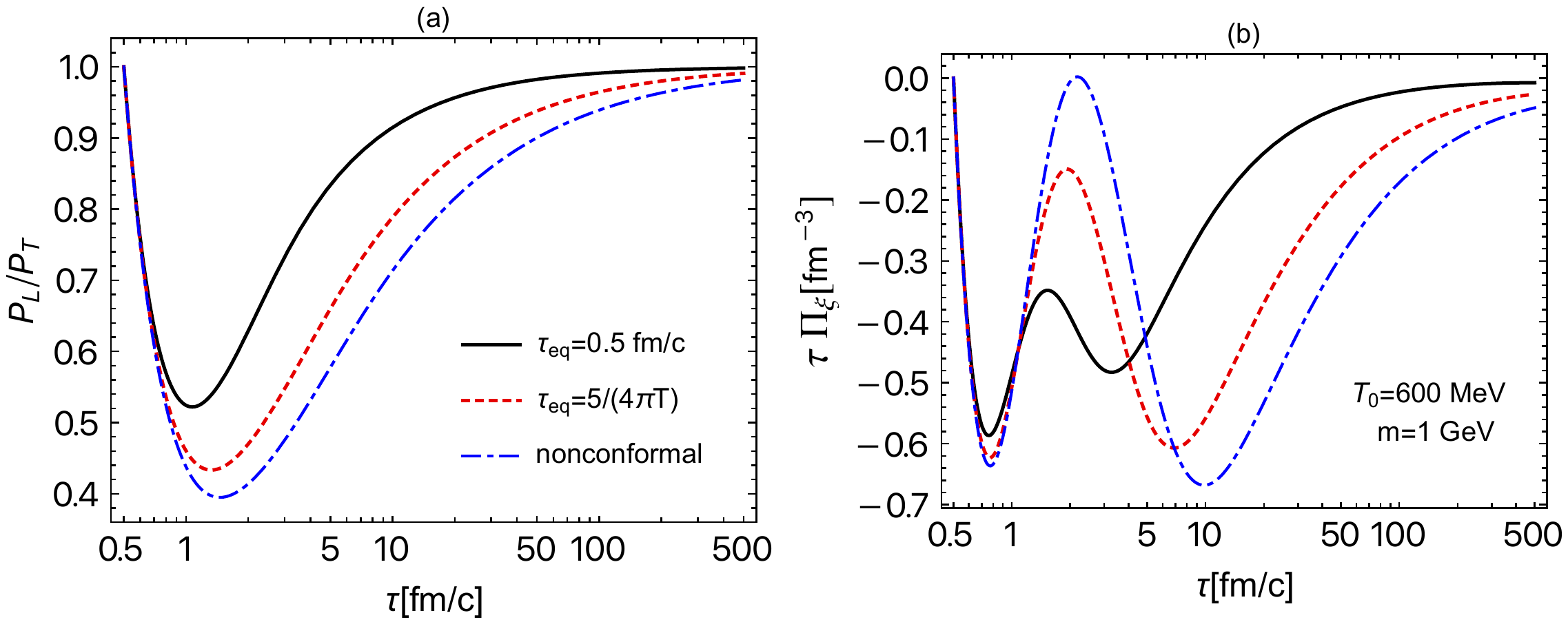}}
\caption{The proper time evolution of the pressure anisotropy and the scaled bulk pressure, left and right panels, respectively. In both panels, the mass is assumed to be constant $m=1$ GeV where different curves correspond to constant $\tau_{\rm eq}$ (black-solid line), conformal  $\tau_{\rm eq}$ (red-dashed line), and nonconformal  $\tau_{\rm eq}$ (blue-dotted-dashed line).  }
\label{fig:bulkVars}
\end{figure}

\begin{figure}[t!]
\centerline{
\hspace{-1.5mm}
\includegraphics[width=1\linewidth]{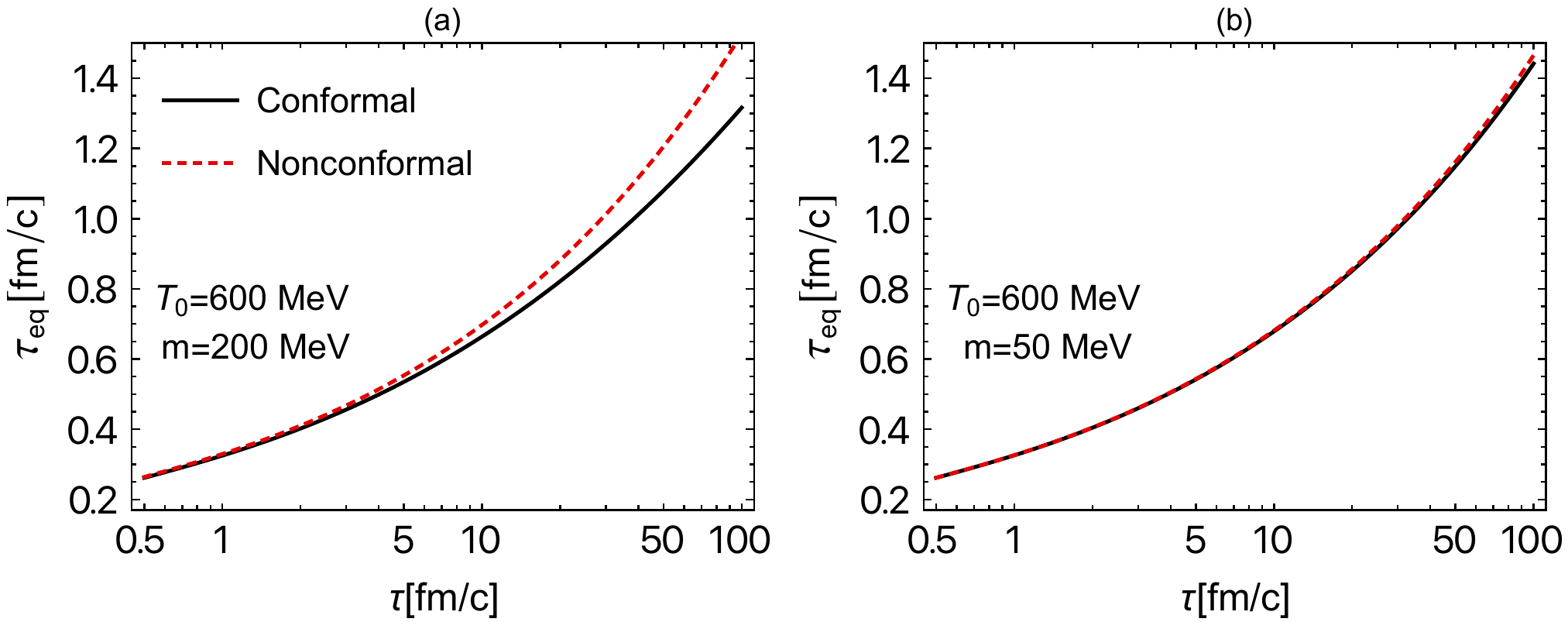}}
\caption{The proper time evolution of  conformal (black-line) and nonconformal (red-dashed line) $\tau_{\rm eq}$ for $m=200$ MeV and $m=50$ MeV, left and right panels respectively.  }
\label{fig:teq}
\end{figure}

\begin{figure}[h!]
\centerline{
\hspace{-1.5mm}
\includegraphics[width=0.9\linewidth]{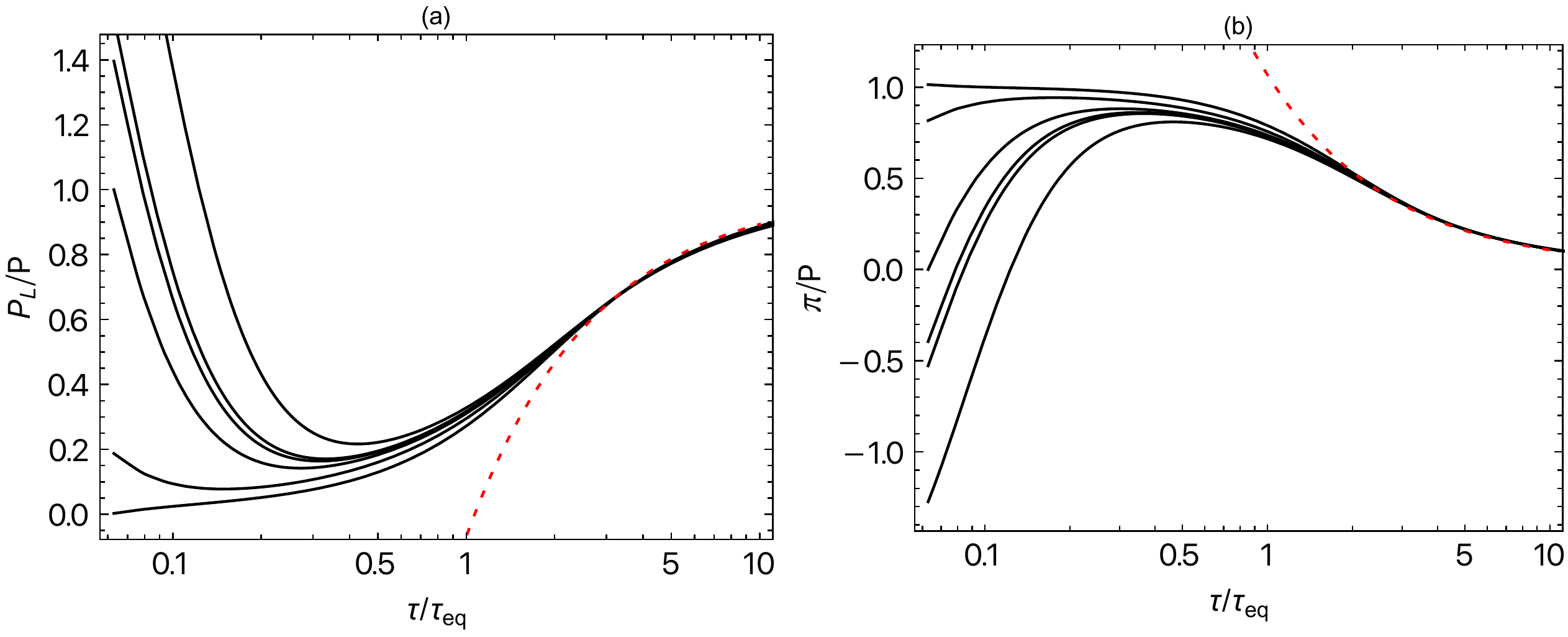}}
\centerline{
\hspace{-1.5mm}
\includegraphics[width=0.9\linewidth]{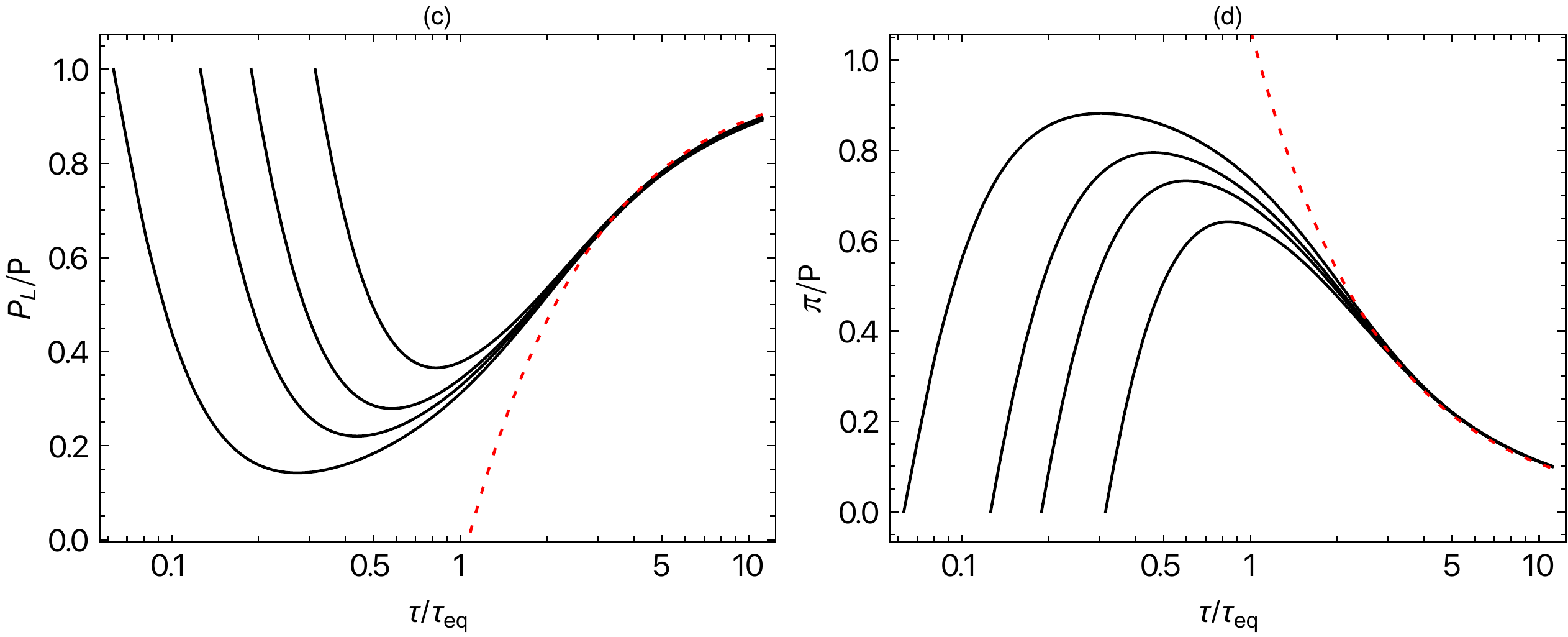}
}
\centerline{
\includegraphics[width=0.9\linewidth]{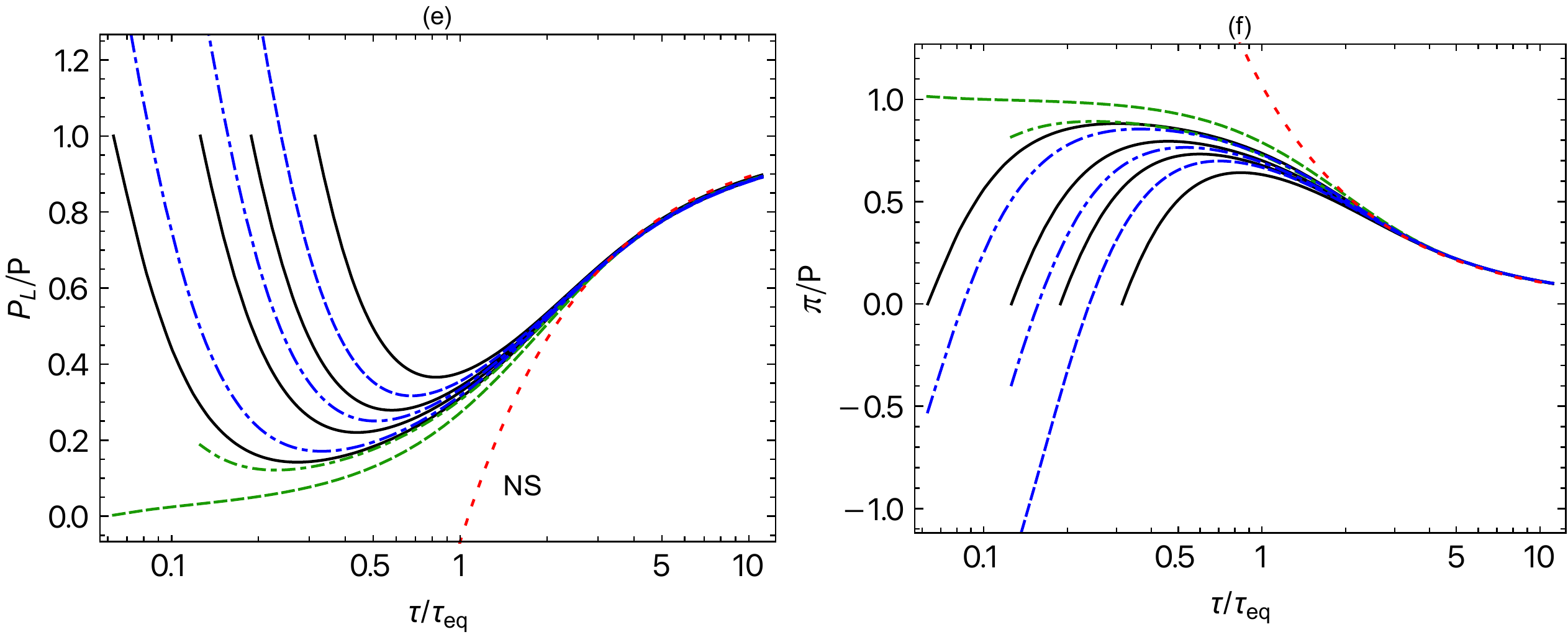}
}
\caption{Scaled time evolution of the scaled longitudinal pressure (left panel) and the scaled shear stress $\pi/p$ (right panel). In the top row, the initial time is assumed fixed at $\tau_i=0.1$ fm/c while anisotropy parameters are different for each curve. In the middle row,  the initial time is different for each curve which is taken to be  \{0.1, 0.2, 0.3, 0.5\} fm/c while anisotropy parameters are held similar. In the bottom row, initial times and anisotropy parameters are varied. In all panels, the mass is assumed to be constant and taken here to be $m=200$ MeV.}
\label{fig:constmass}
\end{figure}

\section{Results}
\label{sec:results}

In this section, we present our findings regarding the existence of early-time attractors in the nonconformal Bjorken (0+1D) expansion using anisotropic hydrodynamics. We will consider two cases where the mass is constant first and then when the mass is thermal $m(T)$. In each case, we will study the behavior of the scaled longitudinal pressure $ {\cal P}_L/{\cal P}$, the scaled shear stress pressure $\pi/P$, and the scaled bulk pressure $\Pi/P$. The shear stress pressure $\pi$ and the bulk pressure $\Pi$ are defined as
\be
\pi \equiv \frac{2}{3} \left({\cal{P}}_{\rm{T}}- {\cal{P}}_{\rm{L}}\right) \, .
 \label{eq:shearpress_defn}
\ee
\be
\Pi \equiv \frac{1}{3} \left({\cal{P}}_{\rm{L}}+ 2 {\cal{P}}_{\rm{T}}\right)-{\cal{P}}_{\rm{eq}} \, .
 \label{eq:bulkpress_defn}
\ee
%
\subsection{Case I: Nonconformal attractors with a constant mass}
\label{subsec:results1}
First, we consider the mass to be constant throughout the evolution of the system. In this case, the dynamical equations of 0+1D anisotropic hydrodynamics were already derived in Ref.~\cite{1405.1355}. They also could be derived from the final equations listed in Sec.~\ref{sec:aHydroQP} as explained there. For the purpose of testing the code written for this case, In Fig.~\ref{fig:bulkVars}, we show the proper time evolution of the pressure anisotropy and the scaled bulk pressure $\tau \Pi$, left and right panels respectively. In each panel, the different curves correspond to constant $\tau_{\rm eq}$ (black-solid line), conformal  $\tau_{\rm eq}$ (red-dashed line), and nonconformal  $\tau_{\rm eq}$ (blue-dotted-dashed line). The constant $\tau_{\rm eq}=0.5$ fm/c case is what the authors of~\cite{1405.1355} used, while the conformal ansatz $\tau_{\rm eq}=5\bar{\eta}/T$ is used in other works as an approximation, e.g. Ref.~\cite{2107.05500,2107.10248}.  However, in this work, we will use the full nonconformal $\tau_{\rm eq}$~\cite{Czyz:1986mr} which has been used in many 3+1D phenomenological comparisons (see e.g.~\cite{1703.05808,1807.04337,2007.03939,2008.07657}). From the black line, we were able to reproduce the results shown in~\cite{1405.1355} for these two observables. From the other two curves, the conformal and nonconformal ansatzes of the $\tau_{\rm eq}$, one can see that there are differences between them, especially at intermediate times, but eventually, they both approach the asymptotic behavior of each observable at extremely very late times. In all three cases, all curves at very late times approach unity for the pressure anisotropy and zero for the scaled bulk pressure as one may expect from the approximate isotropization.


To investigate the differences between the conformal and nonconformal ansatzes more, we plot in Fig.~\ref{fig:teq} the temporal dependence of $\tau_{\rm eq}$ for both ansatzes at different masses, $200$ MeV and $50$ MeV. We see that the conformal ansatz is a good approximation for small masses, especially at early times when $m/T < 1$. This means that the system behaves conformally at very early times, but the nonconformal corrections become important at later times due to the decrease of the temperature while $m$ is constant. In all figures below, we use the nonconformal $\tau_{\rm eq}$.

\begin{figure}[h!]
\centerline{
\hspace{-1.5mm}
\includegraphics[width=0.88\linewidth]{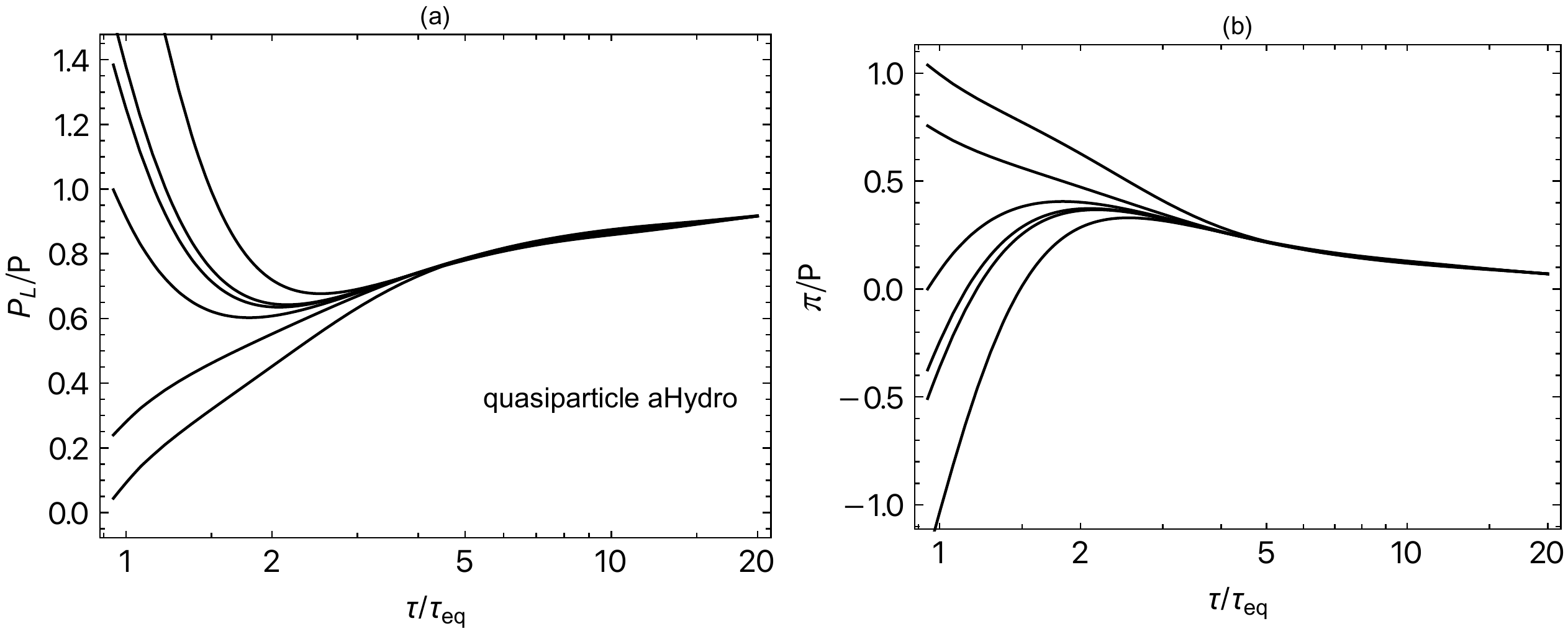}}
\centerline{
\hspace{-1.5mm}
\includegraphics[width=0.88\linewidth]{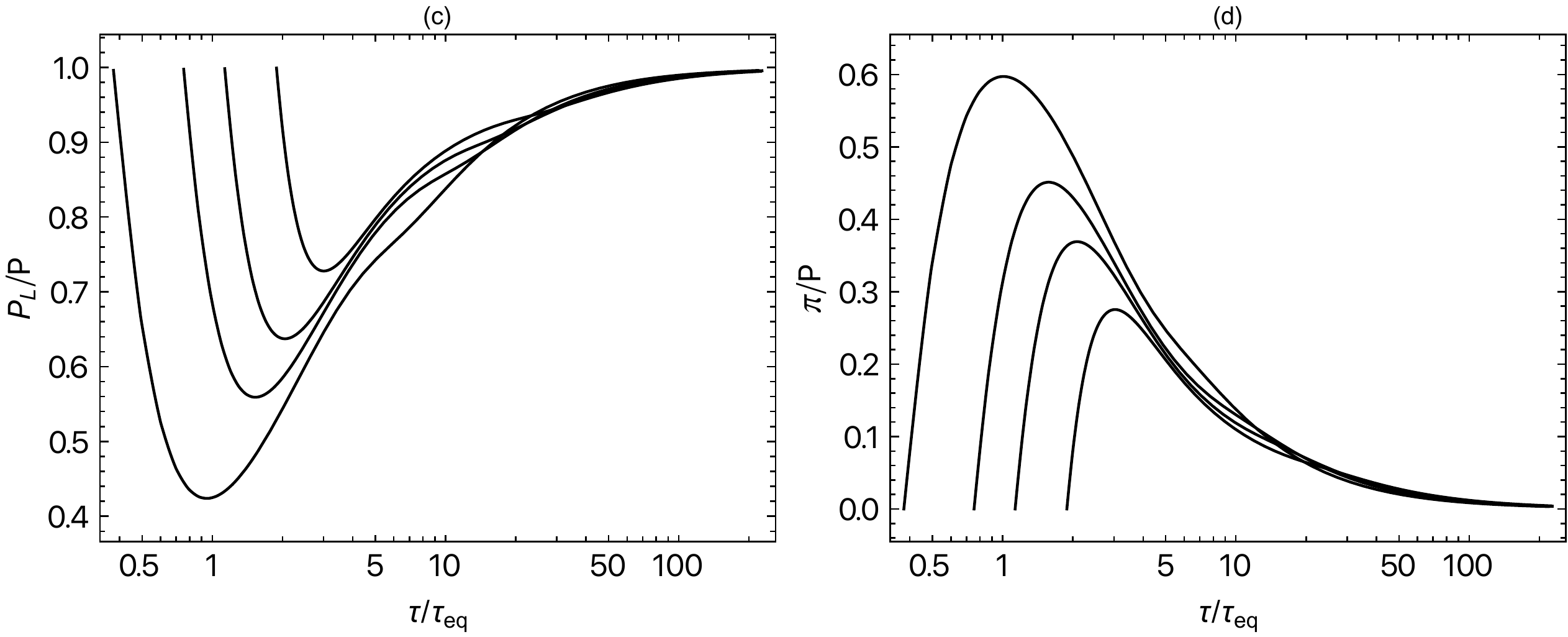}
}
\centerline{
\includegraphics[width=0.88\linewidth]{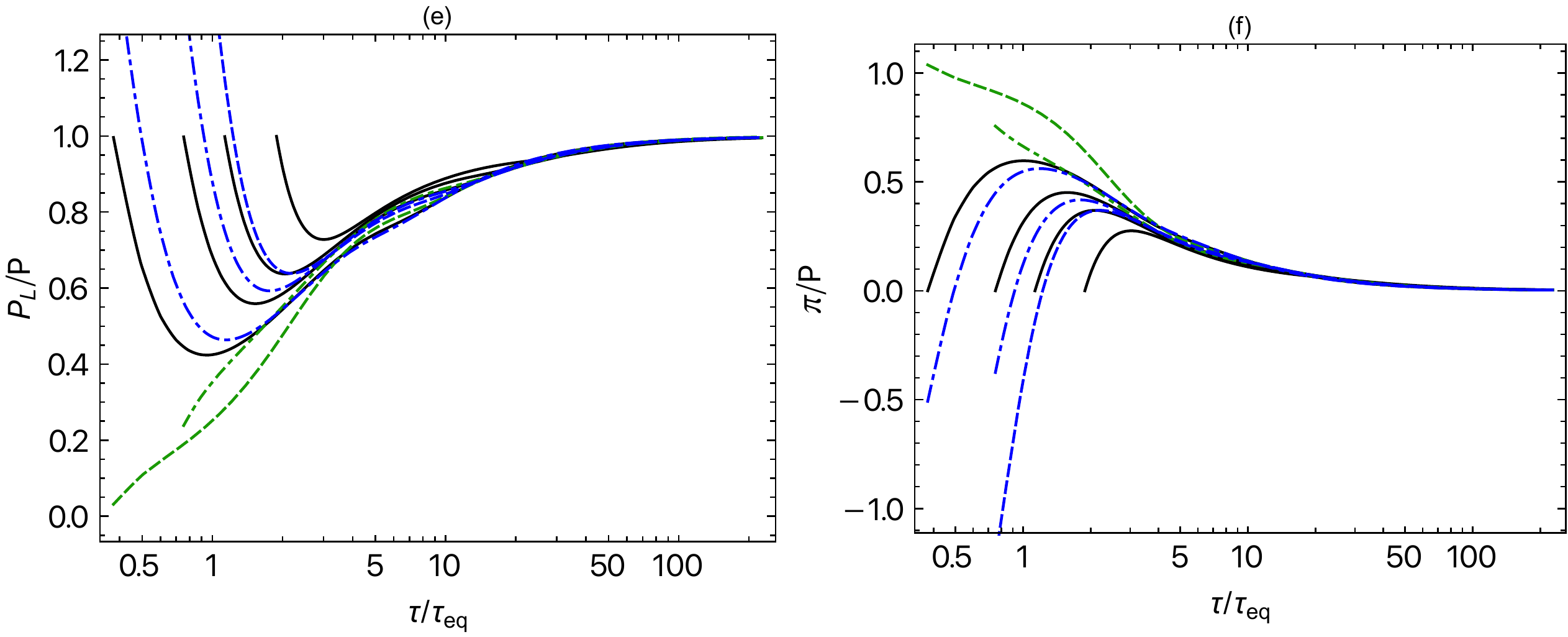}
}
\caption{Scaled time evolution of scaled the longitudinal pressure (left panel) and the scaled shear stress $\pi/P$ (right panel) using the quasiparticle anisotropic hydrodynamics approach. In the top row, the initial time is taken to be $\tau_i=0.25$ fm/c while the anisotropy parameters are varied for each curve~\cite{2203.14968}. In the middle row,  the initial time is varied which is taken to be  \{0.1, 0.2, 0.3, 0.5\} fm/c while the anisotropy parameters are similar for each curve. In the bottom row,  initial times and/or anisotropy parameters are varied for each curve. In all panels, the mass is assumed to be temperature dependent $m(T)$ which is obtained by tuning to the EoS from lattice QCD calculations~\cite{1509.02913,1007.2580}.}
\label{fig:qp}
\end{figure}

For coming comparisons, we list the results of the hydrodynamic attractor given by the conformal Navier-Stokes theory (detailed derivation can be found in Ref.~\cite{1709.06644}). The shear stress pressure attractor, in this case, is given
\be
\frac{\pi }{\epsilon} = 4(\phi-\frac{2}{3}) \, .
 \label{eq:}
\ee
Which is upon using the conformal equation of state $\epsilon=3P$ gives
\be
\bar{\pi}=\frac{\pi }{P} = 12(\phi-\frac{2}{3}) \, .
 \label{eq:}
\ee
On the other hand, the scaled longitudinal pressure is $\bar{{\cal P}}_L =1-\bar{\pi}$.

The Navier-Stokes attractor solution $\phi(\bar{w})$ is given by
\be
\phi = \frac{2}{3}+\frac{4}{9}\frac{c_{\eta/\pi}}{\bar{w}} \, .
 \label{eq:}
\ee
where $c_{\eta/\pi}=1/5$ and $\bar{w}=\tau/\tau_{\rm eq}$~\cite{1709.06644}.

First, we investigate the existence of the early-time (pullback) attractor and the late-time (hydrodynamic) attractor in the case of a constant mass.  In Fig.~\ref{fig:constmass}, we show the scaled time evolution ($\bar{\tau}=\tau/\tau_{\rm eq}$) of the scaled longitudinal pressure and the scaled shear stress $\pi/P$, left and right columns, respectively. In the evolution, the shear viscosity to entropy density ratio is assumed to be $4 \pi\eta/s=10$, the initial temperature is $T_0=500$ MeV and the mass is assumed to be constant $m=200$ MeV. In the top row, the initial time is assumed fixed at $\tau_i=0.1$ fm/c, while the anisotropy parameters are different for each curve. In the middle row,  the initial time is different for each curve while the anisotropy parameters are held similar. The initial times considered here are \{0.1, 0.2, 0.3, 0.5\} fm/c which we picked arbitrarily. In the bottom row, initial times and anisotropy parameters are varied. In the left column, the red long-dashed line is the conformal (i.e. bulk corrections are ignored) Navier-Stokes results~\cite{1709.06644}. 

As can be seen from this figure,  at early times, trajectories are strongly separated due to their strong dependence on the initial conditions.  This dependence in the case of $P_L/P$  is washed out in a small-scaled time $\sim 1$, where they emerge with the late-time Navier-Stokes attractor at a later time and become indistinguishable. At very late times, one can still see slight differences between the curves, which are purely numerical in origin. A different scaling can cancel out these differences, i.e. $ {\cal P}_L/{\cal P}_T$ or as shown in the right panel when considering the shear stress when they also got canceled out by the subtraction. 

On the other hand, the scaled shear stress $\pi/P$ does not show an early-time collapse of the solutions. As seen from the middle row, in the case of $P_L/P$, all solutions collapse before emerging with the NS attractor. However, in the case of $\pi/P$, they fully collapse almost at the same time they emerge with the NS solution.  This means that the real attractor is $ {\cal P}_L/{\cal P}$ as pointed out in~\cite{2107.05500,2107.10248,2210.00658} using kinetic theory. Moreover, for the scaled bulk pressure, there is no early-time attractor. We note that the bulk pressure corrections are very small, which could explain why $\pi/P$ curves look like they may represent an early-time attractor. To allow for large negative bulk viscous pressures and to explore the allowed regions of initial conditions fully, the authors of~\cite{2107.10248}  introduced a modified form of aHydro with a fugacity parameter allowing for control of the magnitude of the distribution function.  More comparisons to this method and exact kinetic theory solutions are left for future work.

  In all panels of Fig.~\ref{fig:constmass}, we take $m=200$ MeV; however, a similar conclusion can be drawn when other masses are used (not shown here). We note that similar results for the left panel of the top row are obtained in~\cite{2107.10248} using a modified anisotropic hydrodynamics approach. 
\subsection{Case II: Nonconformal attractors with a thermal mass}
\label{subsec:results2}

\begin{figure}[t!]
\centerline{
\hspace{-1.5mm}
\includegraphics[width=0.49\linewidth]{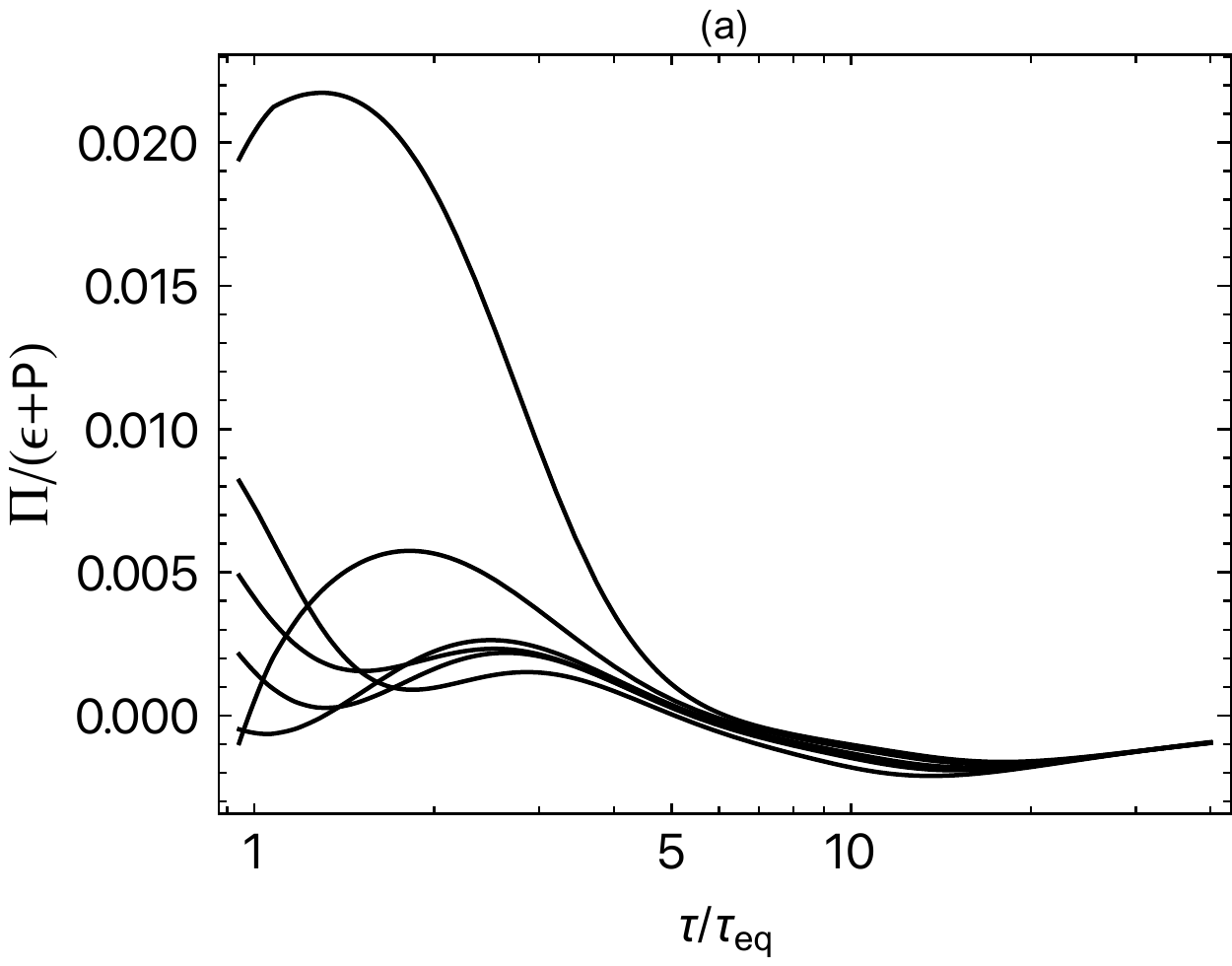}
\hspace{1.5mm}
\includegraphics[width=0.49\linewidth]{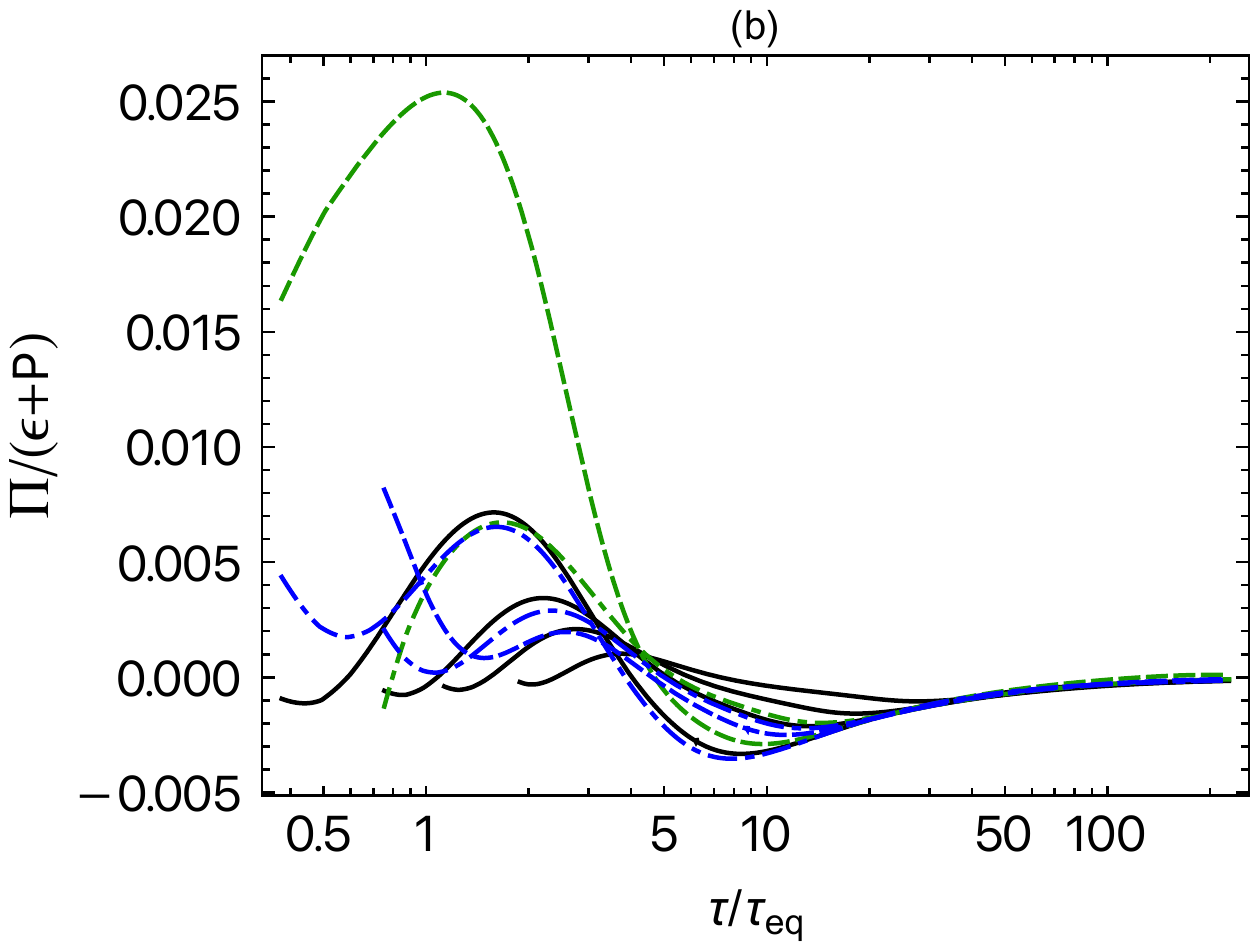}}
\caption{The scaled time evolution of the scaled bulk pressure using the aHydrQP approach where the mass is assumed to be temperature dependent $m(T)$ obtained by tuning to the EoS from lattice QCD calculations~\cite{1509.02913,1007.2580}. The left panel shows the evolution when the initial time is fixed for all curves~\cite{2203.14968}, while the right panel shows the evolution for different initial times and anisotropy parameters. }
\label{fig:qp-bulk}
\end{figure}

All results discussed so far are restricted to the case of a constant mass. Next, we consider the case when a realistic equation is used with a single thermal mass $m(T)$. Numerical solutions of Eqs.~(\ref{eq:final1-thermal}-\ref{eq:final-matching-thermal})  accompanied with a realistic equation of state are solved using Mathematica and were developed in prior works~\cite{1509.02913,2203.14968}. The code is modified here to produce the desirable bulk observables to examine the existence of early-time attractors. Using this model,  results are shown In Fig.~\ref{fig:qp}, which is similar to  Fig.~\ref{fig:constmass}, except the fact that we are using the quasiparticle anisotropic approach with $4 \pi\eta/s=2$ and $T_0=600$ MeV. From the top row, we see the existence of a very late-time universal attractor  ($\bar{\tau}\sim 5$) which is in agreement with results shown in~\cite{2203.14968}. However, when initial times and/or anisotropy parameters are varied, as shown in the middle and bottom rows, the solutions converge more slowly with no strict universal attractor. As a result, this late-time universality is not completely broken, but there is a band instead of a smooth line as shown in Fig.~\ref{fig:constmass} for the constant mass. For example, in the middle row, the $ {\cal P}_L/{\cal P}_T$ solutions at  ($\bar{\tau}\sim 5$) differs by $\sim 10$\%. Finally, in Fig.~\ref{fig:qp-bulk}, we show the evolution of the scaled bulk pressure as a function of $\bar{\tau}$ where we observe a complete absence of the early-time attractor. Similar observations, not shown here, are seen in the case of a constant mass. A final note on the numerical results, as mentioned above, the Mathematica code written for the quasiparticle anisotropic hydrodynamics has been used and tested before in Refs.~\cite{1509.02913,2203.14968}.  We also have tested the code this time in different possible ways, and we believe that the absence of the strict attractor is a nontrivial effect caused by the temperature dependence of the mass.

 Finally, we note that we have no direct explanation of why only a semi-universality of the attractor exists in the case of aHydro when a realistic equation is used. The exact solutions of the kinetic theory for thermal masses may shed more light on these observations. To the best of our knowledge, there have been no studies of exact solutions of the Boltzmann equation where a thermal mass is included in the dynamics.

\section{Conclusions and Outlook}
\label{sec:conclusions}
 In this work, we examined the existence of early-time (pull-back) and late-time (hydrodynamic) attractors of systems undergoing Bjorken expansion using anisotropic hydrodynamics. We first assumed a single constant mass in the dynamics and studied the evolution of the pressure anisotropy and scaled shear stress tensor.  We found the evolution is insensitive to variations of initial conditions for the scaled longitudinal pressure converging onto an early-time universal curve, which eventually merges with the late-time Navier-Stokes attractor. These findings agree with observations from kinetic theory's exact results shown in~\cite{2107.10248,2210.00658}.
 
We then considered the case where the mass is assumed to be thermal $m(T)$ with a realistic equation of state (lattice QCD) as an input to the dynamics. We demonstrated the absence of strict early-time universal attractors of the scaled longitudinal pressure and shear stress tensor. However, a quasi-universal feature of the evolution at late times remains. This means that the system keeps some memory of the initial conditions where their effects are not completely washed out as found in other approaches. We finally note that such spread of the solutions is rather small, e.g., for the scaled longitudinal pressure solutions at  $\bar{\tau}=5 $, the spread is in the order of $\sim 10$\% for the choices considered in the middle row of Fig.~\ref{fig:qp}

Looking to the future, it is important to find the exact solutions of kinetic theory in the case of quasiparticles. It will not be an easy modification for the case of constant mass, but we believe it is doable. If found, they may help in fully understanding the existence of quasi-universality of the late-time attractors. Moreover, we plan to compare the modified anisotropic hydrodynamics approach introduced in Ref.~\cite{2107.10248} with the standard anisotropic hydrodynamics approach used in this work and compare both to exact kinetic theory results using constant mass. Another interesting project is to study the far-from-equilibrium attractors with a nonvanishing baryon chemical potential as has been done in other frameworks~\cite{2007.15083,2207.04086,2209.10483} using aHydro. These projects are left for future research.

\section*{Acknowledgments}

The author would like to thank M. Strickland and S. Jaiswal for fruitful discussions. The author also acknowledges the support of Imam Abdulrahman Bin Faisal University.


\bibliography{RealisticAttractor}

\end{document}